\documentclass[sigconf]{acmart}
\AtBeginDocument{%
  }

\setcopyright{acmlicensed}
\copyrightyear{2018}
\acmYear{2018}
\acmDOI{XXXXXXX.XXXXXXX}
\acmConference[Conference acronym 'XX]{Make sure to enter the correct
  conference title from your rights confirmation email}{June 03--05,
  2018}{Woodstock, NY}
\acmISBN{978-1-4503-XXXX-X/2018/06}

\usepackage{multirow}   
\usepackage{makecell}   
\usepackage{graphicx}   
\usepackage{threeparttable} 

\begin{document}

\title{Causal Inspired Multi Modal Recommendation}

\author{Jie Yang}
\affiliation{%
  \institution{National University of Singapore
Master of Industrial and Systems Engineering}
  \country{Singapore}}
\email{2454493756@qq.com}

\author{Chenyang Gu}
\affiliation{%
  \institution{East China Normal University
Master of Library and Information Science}
  \city{Shanghai}
  \country{China}}
\email{z13521765427@163.com}

\author{Zixuan Liu}
\affiliation{%
  \institution{Shandong Normal University}
  \city{Jinan}
  \state{Shandong}
  \country{China}}
\email{lzx119229@163.com}

\renewcommand{\shortauthors}{Jie Yang et al.}

\begin{abstract}
Multimodal recommender systems enhance personalized recommendations in e-commerce and online advertising by integrating visual, textual, and user-item interaction data. However, existing methods often overlook two critical biases: (i) modal confounding, where latent factors (e.g., brand style or product category) simultaneously drive multiple modalities and influence user preference, leading to spurious feature-preference associations; (ii) interaction bias, where genuine user preferences are mixed with noise from exposure effects and accidental clicks. To address these challenges, we propose a \textbf{Ca}usal-inspired \textbf{m}ultimodal \textbf{Rec}ommendation framework. Specifically, we introduce a dual-channel cross-modal diffusion module to identify hidden modal confounders, utilize back-door adjustment with hierarchical matching and vector-quantized codebooks to block confounding paths, and apply front-door adjustment combined with causal topology reconstruction to build a deconfounded causal subgraph. Extensive experiments on three real-world e-commerce datasets demonstrate that our method significantly outperforms state-of-the-art baselines while maintaining strong interpretability.
\end{abstract}


\keywords{Do, Not, Use, This, Code, Put, the, Correct, Terms, for,
  Your, Paper}

\received{20 February 2007}
\received[revised]{12 March 2009}
\received[accepted]{5 June 2009}

\maketitle

\section{Introduction}
With the rapid development of online multimedia platforms, multimodal recommender systems (MMRec)\cite{survey} have become increasingly essential in domains such as e-commerce and online advertising, effectively capturing personalized user preferences by integrating item content (e.g., textual and visual descriptions) with historical user–item interactions. Recent advances in deep learning have led multimodal systems to primarily adopt multimodal representation alignment techniques for learning consistent item representations across modalities, while leveraging Graph Neural Networks (GNN)\cite{GNN}, such as LightGCN\cite{LightGCN}, to capture relational structures between users and items.

Despite promising results, current multimodal recommendation approaches often overlook two critical biases (as illustrated in Figure 1). Firstly, \emph{modal confounding} arises as multimodal item features (e.g., images and text) are typically driven by underlying latent factors, such as brand style or product category. Many existing studies ignore this issue, causing models to rely excessively on shallow modality-specific features irrelevant to actual user preferences, thus deteriorating recommendation accuracy and generalization in practical scenarios. Secondly, \emph{interaction bias} exists in user–item interaction graphs, as genuine user preference signals often become entangled with noise introduced by external factors such as exposure bias, popularity effects, and accidental clicks. For instance, popular items may accumulate interactions merely due to high exposure, not because their features match users' genuine interests. Users might also click items out of promotional incentives rather than true interest, leading the recommender to excessively favor popular but irrelevant items.
\begin{figure*}[htbp]
    \centering
    \includegraphics[width=1\textwidth]{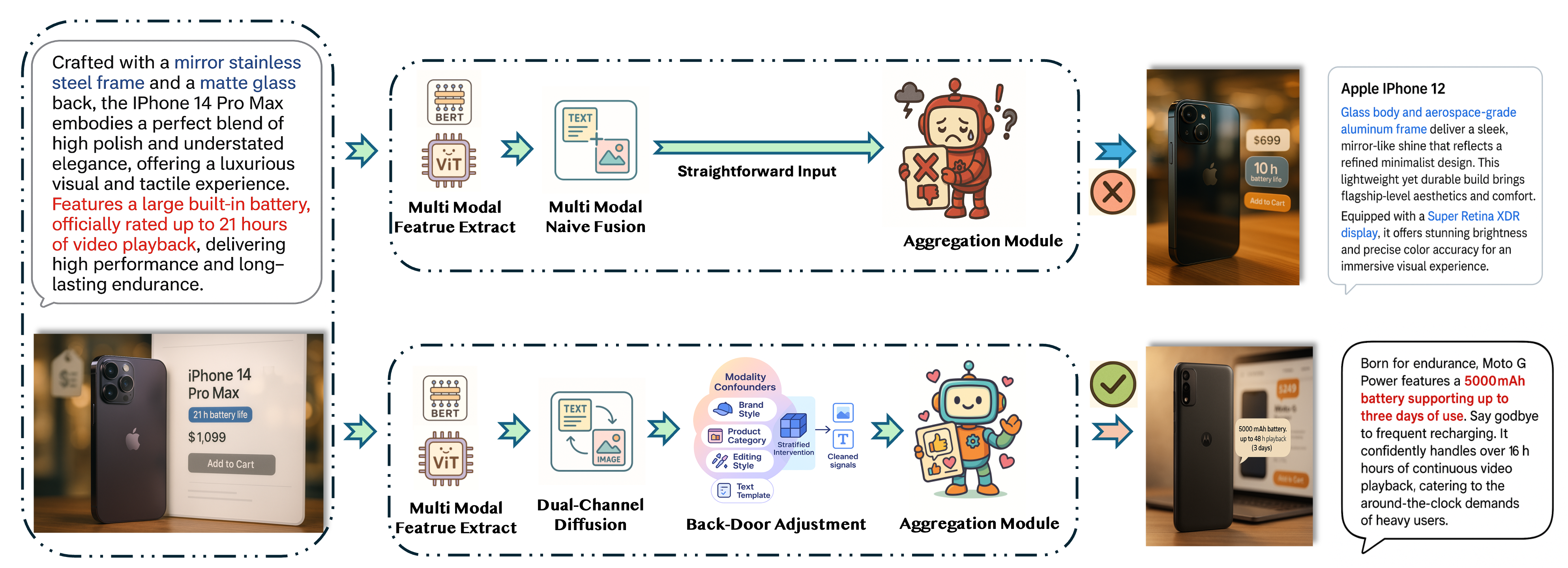} 
    \caption{Comparison of multimodal feature extraction approaches for product descriptions. The top path illustrates a straightforward naive fusion of text and image features, leading to potential misinterpretation due to unaddressed modal confounders. The bottom path shows an enhanced approach using dual-channel diffusion and back-door adjustment to clean and integrate features, resulting in a more accurate representation of the product’s unique qualities, reducing biases from modal confounders.} 
    \label{fig:example_image} 
\end{figure*}
In this paper, we aim to simultaneously tackle modal confounding and interaction bias via causal interventions. We first construct a Structural Causal Model (SCM) to elucidate the multimodal preference generation process. Based on this SCM, we then propose: (1) a dual-channel cross-modal diffusion model to ensure the identifiability of latent modal confounders; (2) a back-door adjustment strategy to conduct causal interventions against modal confounding; and (3) a front-door adjustment strategy to effectively capture genuine causal associations between users and items. Specifically, our contributions are summarized as follows:

\begin{itemize}
    \item We propose a novel causal perspective to decode the underlying mechanisms of user preference formation in multimodal scenarios. Building upon this insight, we develop a framework termed Multimodal Causal Recommendation for improved personalization accuracy.
    
    \item To address modal confounding via back-door adjustment, we introduce a dual-channel diffusion model that identifies hidden modal confounders embedded in multimodal item signals. These confounders are then discretized through a learnable vector-quantized codebook for environment stratification, enabling hierarchical matching between items and modal confounders to block spurious associations effectively.
    
    \item To capture genuine user–item causal associations via front-door adjustment, we note that extracting preference signals from historical interactions (interaction contexts) may include spurious causations. However, directly stratifying these contexts is computationally expensive, rendering back-door adjustments impractical. Thus, we introduce a proxy to simulate interaction contexts based on modal features, leveraging causal topology reconstruction to model actual causal relationships.
    
    \item Empirically, we conduct extensive experiments on three real-world e-commerce datasets. Our results demonstrate that our proposed method consistently outperforms state-of-the-art baselines while providing clear interpretability.
\end{itemize}

\section{Related Work}
\textbf{Multimodal Recommendation.} 
Multimodal recommendation has attracted extensive attention in recent years, with a growing body of research addressing various aspects of the field. Leveraging graph neural networks (GNNs), multimodal graph neural networks such as LGMRec\cite{GNN} and FREEDOM\cite{zhou2023freedom} have been developed to model users' multimodal preferences. Integrating multimodal alignment algorithms, including multimodal contrastive learning (e.g., BM3 \cite{zhou2023bm3}, FETTLE\cite{zhang2024fettle}, AlignRec\cite{1}) and multimodal diffusion models \cite{diffmm} (e.g., Dif, MCDRec), researchers have successfully captured cross-modal representation consistency for items and multimodal relationships between users and items. Following these developments, approaches like MENTOR\cite{xu2025mentor} and MMGCL\cite{mmgcl2022} have employed random graph augmentation  strategies to alleviate biases inherent in interaction data. However, these methods largely overlook biases arising from modal confounding and fail to address interaction biases from the perspective of modal representations.

\textbf{Causal Inference for Recommendation.} 
Causal inference~\cite{pearl2009causality} aims to investigate causal relationships among variables, facilitating robust and stable learning and inference. Recently, integrating deep learning techniques with causal inference has demonstrated significant potential, particularly in computer vision, natural language processing, and recommender systems. However, causal inference remains relatively under-explored in multimodal recommendation scenarios. Existing approaches, such as graph-based causal models, primarily target ID-based recommendation tasks by debiasing interaction data for unbiased recommendations. In multimodal contexts, causal inference has generally been used for tasks such as resolving confounding effects across modalities. Nevertheless, these approaches face challenges in multimodal recommendation scenarios because graph-based methods cannot adequately resolve modal confounding biases, while existing multimodal causal models fail to fully satisfy the requirements of recommendation tasks. In this study, we adopt a causal perspective to decode the user preference generation process in multimodal scenarios and utilize causal intervention techniques to mitigate confounding effects arising from modal confounding and interaction data biases in multimodal recommendation.

\label{gen_inst}

\section{Causal Interpretation }
In this section, we first introduce the notation and formalize the multimodal recommendation task. We then use a Structural Causal Model (SCM) to intuitively analyze how multimodal user preferences are generated. Finally, we discuss the data biases inherent in multimodal recommendations from a causal perspective, employing back-door and front-door adjustments to mitigate these biases.

\subsection{Problem Statement}
Let the set of users be denoted as $U = \{u_1, u_2, u_3\}$, and the set of items as $I = \{i_1, i_2, i_3\}$. Each modality-specific feature input is defined as:
$$E^m = \{E_u^{m} \| E_i^m\} \in \mathbb{R}^{d_m \times (|U|+|I|)},$$
where $m \in M$ denotes modality type ($M$ being the modality set), $d_m$ represents the dimensionality of modality features, and $\|$ indicates the concatenation operation. In this paper, we specifically consider visual modality $V$ and textual modality $T$, i.e., $M = \{v, t\}$, though the proposed framework can extend to additional modalities.

\subsection{A Causal Interpretation for Multimodal Recommendation}
To better understand the data biases within multimodal recommendation processes, we establish an SCM that illustrates causal relationships among five variables: modal confounder $C$, user–item interaction structure $G$, item textual features $T$, item visual features $V$, and true user preference $Y$. Arrows between variables represent causal directions. For simplicity, we assume independence between confounder $C$ and interaction structure $G$. To streamline notations, we denote visual and textual features uniformly as $M$. Under this assumption, the causal relationships can be represented as:

\begin{equation}
\begin{split}
P(V,Y|G,C) &= P(V|G,C)P(Y|V,E,C) \\
P(T,Y|G,C) &= P(T|G,C)P(Y|T,E,C)
\end{split}
\end{equation}

We elaborate on these causal relationships as follows:

\textbf{(1) $V \leftarrow C \rightarrow T$ and $C \rightarrow Y$.} 
The modal confounder $C$ is an intrinsic driver behind multimodal feature generation, existing independently as an exogenous variable influencing both visual ($V$) and textual ($T$) features. Formally, this relationship can be expressed as:
\[
P(V,T|C) = P(V|C)P(T|C).
\]
For instance, a minimalist brand style ($C$) often dictates that product images ($V$) have clean and simple compositions, while textual descriptions ($T$) maintain a concise style. Notably, this common confounder also directly influences user preferences ($Y$), as users are frequently attracted by a coherent overall style.

\textbf{(2) $V \leftarrow G \rightarrow T$ and $G \rightarrow Y$.} 
In recommender systems, the user–item interaction graph $G$ impacts feature extraction for $V$ and $T$ in a data-driven manner, simultaneously shaping the model’s understanding of user preferences $Y$. However, this impact includes both genuine causal effects (e.g., actual user preferences reflected in interactions) and spurious correlations (e.g., inflated popularity due to high exposure). As highlighted by prior studies, relying solely on co-occurrence patterns in $G$ could introduce false associations lacking genuine causal validity.

\textbf{(3) $V \rightarrow Y$ and $T \rightarrow Y$.} 
These direct associations are desirable, as they leverage item modality signals to predict user preferences accurately.

\subsection{Confounders and Causal Treatments}
By analyzing the SCM, we identify two back-door paths between multimodal features and user preferences—taking visual modality as an example: $V \leftarrow C \rightarrow Y$ and $V \leftarrow G \rightarrow Y$, with the modal confounder $C$ and user interactions $G$ acting as confounding variables. This implies certain features used to predict user preference $Y$ are substantially influenced by $C$ and $G$. To eliminate the negative impact of these confounders, we adopt causal inference tools and perform do-calculus on $V$ and $T$ to estimate $P(Y|do(V))$ and $P(Y|do(T))$, where $do(\cdot)$ denotes an intervention operation.  

Specifically, for modal confounding $C$, we employ back-door adjustment techniques to block back-door paths from $C$ to $X$ (illustrated by dashed red arrows), effectively removing confounding effects by implicit environment stratification. Nevertheless, given the vast number of users and their unique interactions, stratifying interaction contexts directly is computationally prohibitive. To tackle interaction confounding ($G$), we use front-door adjustment, introducing intermediate variables $V^*$ and $T^*$ (red nodes in Figure 2c), to simulate more accurate representations that exclude spurious parts of $G$. Importantly, we avoid applying front-door adjustment to $C$, as it requires that intermediate variables depend solely on causal variables without interference from other confounders; modal confounders in real-world contexts often interact with additional latent factors, complicating isolation. These two strategies effectively mitigate confounding biases from $G$ and $C$ as detailed below:

\textbf{Back-Door Adjustment for $C$.}  
To predict user preferences ($Y$) based on modality features ($V$ or $T$), modal confounding ($C$) must be mitigated. We first assume a simplified directed acyclic graph (DAG), with confounder $C$ as the sole parent node of modality $X$. Back-door adjustment is implemented by stratifying $C$ into discrete sets $C = \{c_i\}_{i=1}^{|C|}$ for estimating $P(Y | do(X))$. As modal confounders are hidden within multimodal data, we propose a dual-channel cross-modal diffusion mechanism to approximate modal confounders, coupled with a vector quantization-based hierarchical matching technique, enabling the enumeration of $c_i$ for approximate optimal scenario selection.

\paragraph{Front-Door Adjustment For $G$} After addressing modal confounding ($C$), our next step is to remove biases from the interaction environment ($G$) via front-door adjustment. As shown in Figure 2c, we introduce instrumental variables $X^*$ between features ($X$) and preference ($Y$) to model true causal associations based on user interactions. The causal effect of $X$ on $Y$ is thus estimated by:
\[
P(Y | do(X)) = \sum_{x^*}\sum_{x'} P(X^*=x^*|X)P(Y|X^*=x^*,X=x')P(X=x'),
\]
according to the do/observe exchange rule. From this, we infer $P(Y | X^*, X)$ through observing $(X, X^*)$ pairs. This front-door adjustment yields robust estimates of the causal effect of $X$ on $Y$, avoiding spurious associations due to $G$.

\begin{figure*}[htbp]
    \centering
    \includegraphics[width=1\textwidth]{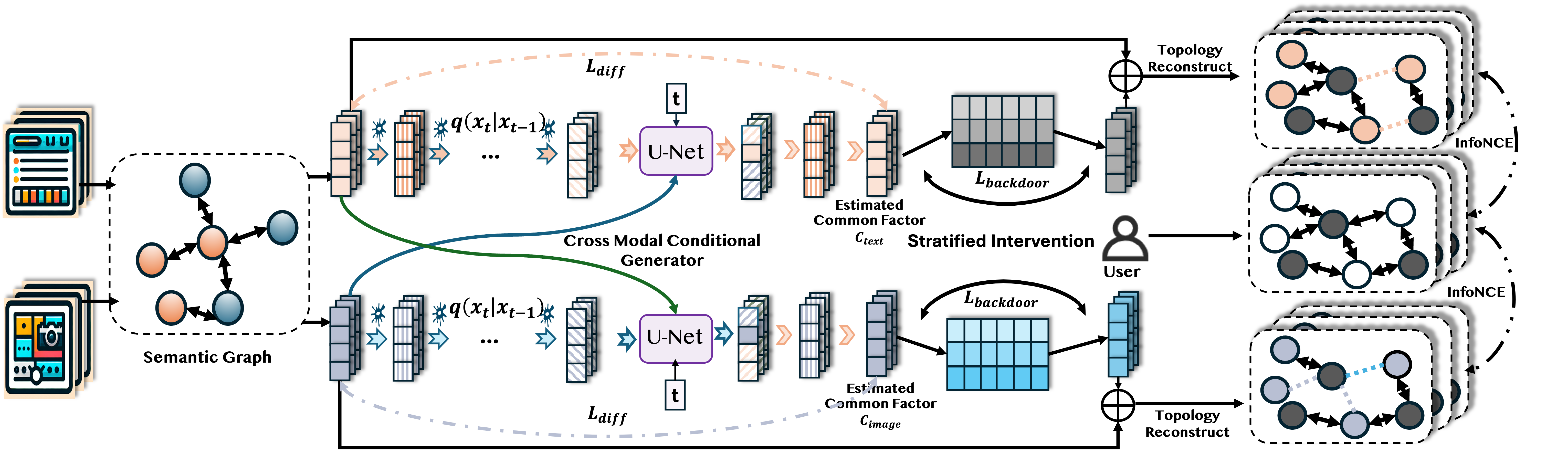} 
    \caption{Overview of the causal-inspired multimodal recommendation framework. The model uses a cross-modal conditional generator to process textual and visual features, incorporating a dual-channel diffusion mechanism to estimate common factors for each modality. The system employs back-door adjustment to eliminate modal confounders, followed by a stratified intervention. Topology reconstruction, powered by InfoNCE loss, ensures that the learned causal relationships between modalities and user preferences remain robust and unbiased.} 
    \label{fig:example_image} 
\end{figure*}

\section{Model Instantiation}
We propose a novel causal framework (illustrated in Figure xx) to implement the aforementioned causal interventions. Our model leverages user historical interactions $G$, item visual features $V$, and textual features $T$ to achieve unbiased prediction of user preferences. The detailed descriptions of the model workflow and core modules are provided below.
\paragraph{Back-door Adjustment} As illustrated in Figure xx (section xx), achieving the back-door adjustment described in Equation xx involves two key steps: (i) identifying modal confounders from multimodal data, and (ii) discretizing these confounders. To address this, we introduce a dual-channel cross-modal diffusion module to model modal confounders and employ a learnable codebook for hierarchical matching and discretization.
\paragraph{Front-door Adjustment} As indicated in Figure xx, acquiring $(X^*)$ and collecting $(X,X^*)$ pairs to achieve Equation xx presents two main challenges: (1) enumerating all historical user–item interaction contexts is computationally expensive, especially for large-scale recommender systems; and (2) quantifying the causal effect of $X$ on $X^*$. To overcome these challenges, we introduce a \emph{topology reconstruction deconfounder}, a neural topology-based module that uses an InfoNCE loss on $(X,X^*)$ pairs to capture causal associations between user preferences and item modalities, effectively identifying genuine user preferences.

\subsection{Dual Channel Cross-Modal Diffusion Module}  
To accurately identify and estimate the latent modal confounder ($C$), we propose a Dual Channel Cross-Modal Diffusion Module. Intuitively, one modality can be considered as a generative condition for another modality, and the shared generative factors between the two modalities correspond to the modal confounder $C$ that we aim to estimate. In this approach, we harness the powerful generative capabilities of diffusion models to explicitly estimate the modal confounder representation by conditioning on the different modalities.

The diffusion model consists of two main stages: forward diffusion and reverse denoising, which we define as follows.

\textbf{Forward Diffusion Process:}  
Let the original feature representation of item $i$ in modality $m \in \{V, T\}$ be $x_i^m$. To improve the estimation of the modal confounder, we draw inspiration from FREEDOM and employ a semantic graph based on modality similarity to map the original representations into a common embedding space:  
\[
h_i^m = ISG(x_i^m)
\]
The details of the Semantic Graph implementation are provided in the appendix.  
Next, we use $h_i^m$ as the initial representation for the diffusion process. The forward diffusion process is defined by gradually adding Gaussian noise at each step:  
\[
q(h_t^m | h_{t-1}^m) = \mathcal{N}(h_t^m; \sqrt{1 - \beta_t}h_{t-1}^m, \beta_t \mathbf{I}), \quad t=1, \dots, T
\]
Here, $\beta_t$ is the diffusion coefficient that controls the intensity of noise added at each step. Using the reparameterization trick, we can compute the diffusion process for any step $t$ as:  
\[
q(h_t^m | h_0^m) = \mathcal{N}(h_t^m; \sqrt{\bar{\alpha}_t}h_0^m, (1-\bar{\alpha}_t)\mathbf{I}), \quad \text{where} \quad \bar{\alpha}_t = \prod_{s=1}^{t}(1-\beta_s)
\]
Intuitively, the forward diffusion process progressively corrupts the modality data, eventually resulting in near pure noise.

\textbf{Cross-modal Conditioned Reverse Process:}  
The reverse process aims to recover the clean modality feature representation by generating it conditionally on another modality, thereby explicitly estimating the modal confounder $C$. Specifically, if we wish to recover the original diffusion representation $h_0^m$ of modality $m$ conditioned on the information $x_i^{m'}$ from modality $m'$, we define the reverse process as follows:  
\[
p_\theta(h_{t-1}^m | h_t^m, x_i^{m'}) = \mathcal{N}(h_{t-1}^m; \mu_\theta(h_t^m,t,x_i^{m'}), \Sigma_\theta(h_t^m,t)\mathbf{I})
\]
The variance term $\Sigma_\theta(h_t^m,t)$ is typically fixed or pre-determined, as in DDPM:  
\[
\Sigma_\theta(h_t^m,t) = \sigma_t^2 \mathbf{I} = \frac{1-\bar{\alpha}_{t-1}}{1-\bar{\alpha}_t} \beta_t \mathbf{I}
\]
The mean term $\mu_\theta(\cdot)$ is predicted by a neural network-based conditional estimator $f_\theta(\cdot)$:  
\[
\mu_\theta(h_t^m, t, x_i^{m'}) = \frac{1}{\sqrt{\alpha_t}}\left(h_t^m - \frac{\beta_t}{\sqrt{1-\bar{\alpha}_t}} f_\theta(h_t^m,t,x_i^{m'}) \right)
\]
Here, $f_\theta(x_t^m,t,h_i^{m'})$ can be a model such as a Transformer or U-Net that captures the modal confounder features under cross-modal conditioning.

The total probability of the reverse diffusion process is given by:  
\[
p_\theta(h_{0:T}^m | x_i^{m'}) = p(h_T^m) \prod_{t=1}^T p_\theta(h_{t-1}^m | h_t^m, x_i^{m'})
\]
Without loss of generality, the Diffusion Model (DM) encourages the posterior distribution $q(h_{t-1} | h_t, h_0)$ to approximate the prior distribution $p_\theta(h_{t-1} | h_t)$ during the reverse process. This optimization objective can be expressed using the Kullback-Leibler divergence:  
\[
L_{vlb} = D_{KL} \left( q(h_{t-1} | h_t, h_0) \parallel p_\theta(h_{t-1} | h_t) \right)
\]

Thanks to the DDPM framework \cite{7}, the optimization of our Dual Channel Cross-Modal Diffusion Module can be reduced to a simple Mean Squared Error (MSE) loss as follows:  
\[
L_{dm} = \mathbb{E}_{h_0^m, h_t^m} \left[ \| h_0^m - f_\theta(h_t^m) \|^2 \right]
\]

Using the above cross-modal diffusion model, we define the modal confounder $H_c$ as the noise-free representation recovered by the reverse process, i.e., $H_c^m = \hat{h}_0^m$.
\subsection{Back-Door and Front-Door Adjustment}

\paragraph{Back-Door Adjustment.} After identifying modal confounders, we adopt an environment codebook $c = \{c_1, c_2, \ldots, c_K\}$, inspired by CaST, to discretize modal confounders. This learnable codebook defines a latent embedding space $c \in \mathbb{R}^{K \times D}$, where $K$ denotes the number of codebook vectors, and $D$ is the dimensionality of each latent vector $c_i$. As illustrated in Figure xx, given the confounder representation $H_c(i) \in \mathbb{R}^{D}$ for item $i$, we identify the closest codebook vector in $c$ through nearest-neighbor lookup. The posterior categorical distribution for assignment is computed as:
\[
q(z_{ij} = k | H_e(i)) = 
\begin{cases} 
1, & \text{if } k = \arg\min_{j} ||H_e(i) - e_j||^2, \\[6pt]
0, & \text{otherwise}.
\end{cases}
\]

Once obtaining the latent assignments $z \in \mathbb{R}^{N\times K}$, we replace each row of $H_c$ with its corresponding discrete vector in $c$ to yield the final environment representation $\hat{H}_c \in \mathbb{R}^{N\times D}$. Note that during inference, we employ a soft categorical distribution for generalization to unseen modal confounders, representing the extent to which each item’s modal features are influenced by confounders:
\[
\hat{H}_e(i) = \sum_{j=1}^{K} q(z_{ij} | H_e(i)) e_j,
\]
where the soft assignment probability $q(z_{ij} | H_e(i))$ ranges between 0 and 1. Further implementation details can be found in Appendix G.

\paragraph{Front-Door Adjustment.} After addressing modal confounding, we focus on biases from historical interactions. Our goal is to derive surrogate variables $\hat{X}_i$ (i.e., latent variables $V^*$, $T^*$ in Figure 2c) to simulate user preferences based solely on true causal interactions. To effectively learn these surrogates, we must account for higher-order causal structures inherent to user-item interactions. Considering the user-item causal relationships as topological structures, we reconstruct a causal subgraph based on modal features and user preferences. Specifically, direct causal effects are modeled as retained edges $e_{ij} \odot \rho_{ij}^{(l)}$, capturing genuine user preferences, while indirect confounding effects are modeled as masked edges $(1-\rho_{ij}^{(l)})e_{ij}$ to block spurious associations.

Inspired by [CGI], we parameterize edge masking probabilities between modal features and user preferences using an MLP:
\[
\omega_{ij}^{(l)} = \text{MLP}([e_i^{(l)}; e_j^{(l)}]),
\]
where $e_i^{(l)}$ and $e_j^{(l)}$ represent item modal features and user preferences, respectively. Using reparameterization, binary variables $\rho_{ij}^{(l)}$ are relaxed from Bernoulli distributions as:
\[
\rho_{ij}^{(l)} = \sigma((\log \varepsilon - \log(1-\varepsilon) + \omega_{ij}^{(l)})/\tau),
\]
where $\varepsilon \sim \text{Uniform}(0,1)$, $\tau > 0$ is a temperature hyperparameter, and $\sigma(\cdot)$ denotes the sigmoid function. This yields the causal subgraph:
\[
G_{Causal}^{(l)} = \{N,\{e_{ij} \odot \rho_{ij}^{(l)} | e_{ij}\in E_G\}\}.
\]
Here, $\rho_{ij}^{(l)}\sim\text{Bernoulli}(\omega_{ij}^{(l)})$ indicates the existence of edge $e_{ij}$.

\section{Experiment}
\subsection{Performance Comparison with State-of-the-Art Methods}
\begin{table*}[t]
  \centering
  \caption{Overall performance on Amazon \textbf{Baby}, \textbf{Sports}, and \textbf{Clothing}. Metrics: Recall@K and NDCG@K ($K \in \{10,20\}$). Bold: best; underline: second-best.}
  \label{tab:main}
  \begin{threeparttable}
  \resizebox{\textwidth}{!}{%
  \begin{tabular}{l l *{4}{c} *{4}{c} *{4}{c}}
  \toprule
  \multicolumn{2}{c}{} &
  \multicolumn{4}{c}{\textbf{Baby}} &
  \multicolumn{4}{c}{\textbf{Sports}} &
  \multicolumn{4}{c}{\textbf{Clothing}} \\
  \cmidrule(lr){3-6} \cmidrule(lr){7-10} \cmidrule(lr){11-14}
  \multicolumn{2}{c}{Model} & R@10 & R@20 & NDCG@10 & NDCG@20 & R@10 & R@20 & NDCG@10 & NDCG@20 & R@10 & R@20 & NDCG@10 & NDCG@20 \\
  \midrule
  \multirow{2}{*}{\makecell[l]{\textbf{CF}}}
  & BPR~\cite{rendle2009bpr}      & 0.0357 & 0.0575 & 0.0192 & 0.0249 & 0.0432 & 0.0653 & 0.0241 & 0.0298 & 0.0206 & 0.0303 & 0.0114 & 0.0138 \\
  & LightGCN~\cite{he2020lightgcn} & 0.0479 & 0.0754 & 0.0257 & 0.0328 & 0.0569 & 0.0864 & 0.0311 & 0.0387 & 0.0361 & 0.0544 & 0.0197 & 0.0243 \\
  \midrule
  \multirow{8}{*}{\makecell[l]{\textbf{2016--2023}}}
  & VBPR~\cite{he2016vbpr}     & 0.0423 & 0.0663 & 0.0223 & 0.0284 & 0.0558 & 0.0856 & 0.0307 & 0.0384 & 0.0281 & 0.0415 & 0.0158 & 0.0192 \\
  & MMGCN~\cite{wei2019mmgcn}  & 0.0421 & 0.0660 & 0.0220 & 0.0282 & 0.0401 & 0.0636 & 0.0209 & 0.0270 & 0.0227 & 0.0361 & 0.0120 & 0.0154 \\
  & GRCN~\cite{wei2020grcn}    & 0.0532 & 0.0824 & 0.0282 & 0.0358 & 0.0599 & 0.0919 & 0.0330 & 0.0413 & 0.0421 & 0.0657 & 0.0224 & 0.0284 \\
  & DualGNN~\cite{wang2023dualgnn} & 0.0513 & 0.0803 & 0.0278 & 0.0352 & 0.0588 & 0.0899 & 0.0324 & 0.0404 & 0.0452 & 0.0675 & 0.0242 & 0.0298 \\
  & SLMRec~\cite{tao2023slmrec}   & 0.0521 & 0.0772 & 0.0289 & 0.0354 & 0.0663 & 0.0990 & 0.0365 & 0.0450 & 0.0442 & 0.0659 & 0.0241 & 0.0241 \\
  & LATTICE~\cite{zhang2021lattice}& 0.0547 & 0.0850 & 0.0292 & 0.0370 & 0.0620 & 0.0953 & 0.0335 & 0.0421 & 0.0492 & 0.0733 & 0.0268 & 0.0330 \\
  & FREEDOM~\cite{zhou2023freedom} &
                \underline{0.0627} & \underline{0.0992} & \underline{0.0330} & \underline{0.0424} &
                \underline{0.0717} & \underline{0.0999} & \underline{0.0385} & \underline{0.0479} &
                \underline{0.0621} & \underline{0.0939} & \underline{0.0339} & \underline{0.0420} \\
  & BM3~\cite{zhou2023bm3}       & 0.0533 & 0.0847 & 0.0291 & 0.0356 & 0.0654 & 0.0981 & 0.0357 & 0.0443 & 0.0434 & 0.0649 & 0.0233 & 0.0298 \\
  \midrule
  \multirow{4}{*}{\makecell[l]{\textbf{2024}}}
  & DiffMM*~\cite{jiang2024diffmm}  & 0.0603 & 0.0975 & 0.0327 & 0.0411 & 0.0699 & 0.1017 & 0.0373 & 0.0458 & 0.0600 & 0.0932 & 0.0316 & 0.0388 \\
  & DA-MRS*~\cite{xv2024damrs} & 0.0648 & 0.1001 & 0.0349 & 0.0439 & 0.0728 & \underline{0.1122} & 0.0399 & \underline{0.0493} & 0.0628 & 0.0962 & 0.0345 & 0.0425 \\
  & M3CSR*~\cite{chen2024m3csr}  & 0.0645 & 0.0999 & 0.0353 & 0.0436 & 0.0722 & 0.1040 & \underline{0.0409} & 0.0471 & 0.0627 & \underline{0.0952} & \underline{0.0346} & \underline{0.0426} \\
  & FETTLE~\cite{zhang2024fettle} & \underline{0.0668} & \underline{0.1023} & \underline{0.0356} & \underline{0.0441} &
               \underline{0.0730} & 0.1115 & 0.0397 & 0.0487 &
               \underline{0.0637} & 0.0963 & 0.0344 & 0.0423 \\
  \midrule
  & \textbf{Ours}   & \textbf{0.0704} & \textbf{0.1068} & \textbf{0.0378} & \textbf{0.0471} &
                      \textbf{0.0786} & \textbf{0.1186} & \textbf{0.0430} & \textbf{0.0531} &
                      \textbf{0.0658} & \textbf{0.0976} & \textbf{0.0358} & \textbf{0.0439} \\
  \bottomrule
  \end{tabular}}
  \begin{tablenotes}\footnotesize
    \item[*] Methods reproduced with authors' official implementations/configs when available.
  \end{tablenotes}
  \end{threeparttable}
\end{table*}

We evaluated \textbf{CaMRec} on the Amazon \textbf{Baby}, \textbf{Sports}, and \textbf{Clothing} datasets using the leave-one-out protocol, reporting \textbf{Recall@10/20} and \textbf{NDCG@10/20}. \textbf{CaMRec} was compared with CF baselines such as BPR \cite{rendle2009bpr} and LightGCN \cite{he2020lightgcn}, and several multimodal methods including VBPR \cite{he2016vbpr}, MMGCN \cite{wei2019mmgcn}, GRCN \cite{wei2020grcn}, DualGNN \cite{wang2023dualgnn}, SLMRec \cite{tao2023slmrec}, LATTICE \cite{zhang2021lattice}, FREEDOM \cite{zhou2023freedom}, BM3 \cite{zhou2023bm3}, and recent works such as DiffMM \cite{jiang2024diffmm}, DA-MRS \cite{xv2024damrs}, M3CSR \cite{chen2024m3csr}, and FETTLE \cite{zhang2024fettle}. As shown in Table \ref{tab:main}, \textbf{CaMRec} outperforms strong baselines across datasets and metrics, with an average improvement of \textbf{5.01\%} over the best baseline, and a maximum gain of \textbf{7.71\%} in NDCG@20 on the \textbf{Sports} dataset. Ablation studies (omitted for space) indicate that removing either the diffusion-based confounder extractor or causal subgraph reconstruction results in up to a 5\% drop in performance, highlighting the importance of both backdoor and frontdoor adjustments. Visualizations of environment clusters and pruned graphs further demonstrate \textbf{CaMRec}'s interpretability.

\subsection{Ablation Study}
To validate the effectiveness of key components in our proposed \textbf{CaMRec} framework, we conduct comprehensive ablation studies on the Amazon \textbf{Baby} dataset. We design three variant models:

\begin{itemize}
    \item \textbf{w/o Backdoor}: Removes the back-door adjustment module, including the dual-channel cross-modal diffusion and environment codebook.
    \item \textbf{w/o Frontdoor}: Removes the front-door adjustment module, including the causal topology reconstruction.
    \item \textbf{w/o DCD}: Removes only the Dual-Channel Diffusion component while retaining the environment codebook.
\end{itemize}

\begin{table}[h]
  \centering
  \caption{Ablation study results on Amazon \textbf{Baby} dataset}
  \label{tab:ablation}
  \begin{tabular}{lcccc}
  \toprule
  \textbf{Variant} & \textbf{R@10} & \textbf{R@20} & \textbf{NDCG@10} & \textbf{NDCG@20} \\
  \midrule
  w/o Backdoor & 0.0644 & 0.0984 & 0.0352 & 0.0439 \\
  w/o Frontdoor & 0.0667 & 0.1018 & 0.0354 & 0.0443 \\
  w/o DCD & 0.0655 & 0.0993 & 0.0356 & 0.0440 \\
  \midrule
  \textbf{Full CaMRec} & \textbf{0.0704} & \textbf{0.1068} & \textbf{0.0378} & \textbf{0.0471} \\
  \bottomrule
  \end{tabular}
\end{table}

As shown in Table \ref{tab:ablation}, we observe that:

\textbf{(1) Both causal adjustments are crucial.} Removing either back-door or front-door adjustment leads to significant performance degradation. The back-door adjustment appears to have a more substantial impact, with performance drops of 8.5\% in R@10 and 7.1\% in NDCG@20 when removed, suggesting that modal confounding poses a more critical challenge in multimodal recommendation.

\textbf{(2) Dual-Channel Diffusion is essential.} The w/o DCD variant shows intermediate performance degradation, indicating that the diffusion-based confounder extraction plays a vital role in accurately identifying modal confounders. The performance drop (7.0\% in R@10) demonstrates that simple environment stratification without proper confounder estimation is insufficient.

\textbf{(3) Component complementarity.} The full model achieves the best performance, confirming that back-door and front-door adjustments address complementary aspects of bias in multimodal recommendation systems.

\begin{acks}
To Robert, for the bagels and explaining CMYK and color spaces.
\end{acks}

\bibliographystyle{ACM-Reference-Format}
\bibliography{sample-base}

\appendix

\end{document}